\documentstyle[12pt]{article}
\renewcommand{\theequation}{\arabic{equation}}
\newcommand{\EQ}{\begin{equation}}
\newcommand{\EN}{\end{equation}}

\newcommand{\bear}{\begin{eqnarray}}
\newcommand{\ear}{\end{eqnarray}}

\begin{document}

\topmargin 0pt
\oddsidemargin 5mm
\newcommand{\NP}[1]{Nucl.\ Phys.\ {\bf #1}}
\newcommand{\PL}[1]{Phys.\ Lett.\ {\bf #1}}
\newcommand{\NC}[1]{Nuovo Cimento {\bf #1}}
\newcommand{\CMP}[1]{Comm.\ Math.\ Phys.\ {\bf #1}}
\newcommand{\PR}[1]{Phys.\ Rev.\ {\bf #1}}
\newcommand{\PRL}[1]{Phys.\ Rev.\ Lett.\ {\bf #1}}
\newcommand{\MPL}[1]{Mod.\ Phys.\ Lett.\ {\bf #1}}
\newcommand{\JETP}[1]{Sov.\ Phys.\ JETP {\bf #1}}
\newcommand{\TMP}[1]{Teor.\ Mat.\ Fiz.\ {\bf #1}}

\renewcommand{\thefootnote}{\fnsymbol{footnote}}

\newpage
\setcounter{page}{0}
\begin{titlepage}
\begin{flushright}
UFSCARF-TH-93-10
\end{flushright}
\vspace{0.5cm}
\begin{center}
{\large  The class of universality of integrable and isotropic GL(N) mixed
magnets}\\
\vspace{1cm}
\vspace{1cm}
{\large S.R. Aladim and  M.J.  Martins } \\
\vspace{1cm}
{\em Universidade Federal de S\~ao Carlos\\
Departamento de F\'isica \\
C.P. 676, 13560~~S\~ao Carlos, Brasil}\\
\end{center}
\vspace{1.2cm}

\begin{abstract}
We discuss a class of transfer matrix
built by a particular combination of isomorphic and non-isomorphic
GL(N) invariant vertex operators. We construct a conformally invariant magnet
constituted of an
alternating mixture of GL(N) ``spins'' operators at different order of
representation. The
corresponding central charge is calculated by analysing the low temperature
behaviour of the
associated free energy. We also comment on possible extensions of our results
for  more general classes of
mixed systems.
\end{abstract}
\vspace{.2cm}
\vspace{.2cm}
\centerline{June 1993}
\end{titlepage}

\renewcommand{\thefootnote}{\arabic{footnote}}
\setcounter{footnote}{0}

\newpage
\section{Introduction}
One of the most useful methods of constructing an integrable one dimensional
quantum spin chain
has been to find solutions of the
Yang-Baxter \cite{YB,YB1} equations. In the context of the lattice models,
such solutions define vertex operators acting on the tensor product of
two vector spaces $V \otimes h_{\alpha}$
of the local states on the horizontal and vertical lines of a given site
$\alpha
   $ of the two dimensional
lattice. The complete Hilbert space of the
model on a lattice of L sites is $\prod_{\alpha=1}^L \otimes h_{\alpha}$.
The vector space $V$ is an auxiliary space which
is useful in the definition of the associated transfer matrix.
Defining $R_{h,\alpha}^V( \mu)$ as such
vertex operator, the corresponding transfer matrix can be expressed by
\cite{YB1
   ,FT}
\EQ
T(\mu)= Tr_{V}[ R_{h,L}^V(\mu) R_{h,L-1}^V(\mu) \cdots R_{h,1}^V(\mu)]
\EN
where the trace is carried out in the auxiliary space $V$ and $ \mu$ is a
variable which parametrizes the
Yang-Baxter solution.

A class of important solutions of the Yang-Baxter equation is that
isomorphic on the horizontal
and vertical spaces~($V \equiv h_{\alpha}$) and
rational on the spectral parameter $\mu$. One example is
the vertex operator composed of generators invariant by some
of the semi-simple $A,D,E$ Lie algebra at certain
order $k$ of its representation. The main feature of
these solutions is that they are believed to define conformally
invariant quantum spin chains \cite{CV,AF,MA} which realize the
class of universality of Wess-Zumino-Witten-Novikov~(WZWN)
\cite {KZ} field theories with topological charge $\tilde{k}$. Recently, de
Vegaand Woynarovich \cite{VW} have
pointed out that other interesting classes of conformally invariant models can
be constructed. These
theories are obtained by combining isomorphic and non-isomorphic vertex
operators invariant by some group
symmetry at different order of representation. As a typical example one can
consider an
alternating combination of isomorphic vertex at even sites and of
non-isomorphic
    operator at odd sites.
Considering that the isomorphic~(non-isomorphic) vertex operator
acts on the vector space $V^{(k)} \otimes h_{\alpha}^{(k)}$
$(V^{(k)} \otimes h_{\alpha}^{(k^{'})})$ of order $k$ and $k^{'}$,
the composed theory described above will lead to a mixed magnet chain
with ``spins'' operators of order $k$ and $k^{'}$. Following this approach, de
Vega and Woynorovich \cite{VW}
have constructed  an alternating anisotropic
Heisenberg chain of spins $1/2$ and $1$ and analysed its behaviour
in the thermodynamic limit. More recently, the central charge of the
conformally
    invariant
isotropic $SU(2)$ mixed chain has been computed independently in refs.
\cite{VN,
   AM} for spins 1/2-1 and
1/2-S, respectively.

Its seems interesting to investigate the critical properties of conformally
invariant mixed spins chain
for a more general class of group symmetry
and its possible relations with WZWN field theories. In this
paper we  study an isotropic alternating $GL(N)$ model in which the
isomorphic vertex is in the fundamental
representation~($k=1$) and the non-isomorphic operator is defined in the tensor
   space product of the
fundamental and the order $k$ of representations. By using the thermodynamic
Bethe ansatz, we have calculated
the associated central charge in the case of a conformally invariant mixed
$GL(N
   )$ system. It is
noted that this central charge can be decomposed in terms of the conformal
anomaly of two WZWN theories
with different topological charges $\tilde{k}=1$ and $\tilde{k}=k-1$,
respectively. We also discuss the generalization
of our results in the case of semi-simple Lie algebras at arbitrary symmetric
representation.

This paper is organized as follows. In sect.2 we define the $GL(N)$ mixed
magnetand we discuss its
diagonalization by the quantum inverse scattering~(QIS) formalism. In sect.3 we
   use the thermodynamic
Bethe ansatz~(TBA) approach in order to investigate the low temperature
behaviour of a conformally
invariant mixed $GL(N)$ system. Sect.4 is devoted to our discussion on possible
   extensions of the
results of sect.3 to other Lie algebras. Sect.5 contains our conclusions. In
Appendices A and B
we summarize some details of the QIS approach and we present extra numerical
checks of the finite
size behaviour for the ground state energy, respectively.

\section{The mixed GL(N) integrable model}

A rational $GL(N)$ invariant vertex operator has been found by Kulish,
Reshetikhin and Sklyanin \cite{KRS} on
their study of group invariant solutions of the Yang-Baxter relation. The
$GL(N)
   $
non-isomorphic vertex defined on
the fundamental and on the symmetric order $k$ of representation is given by
the
    expression \cite{KRS}
\EQ
R_{k,\alpha}^1 (\mu) = \mu -(k-1)/2 +\sum_{i,j} e^{ij} \otimes E^{ji}_{\alpha}
\label{R1k}
\EN
where $e^{ij}~(E^{ij})$ are the generators of $GL(N)$ in the fundamental~(order
   $k$) representation. For
the fundamental representation the matrix elements of $e^{ij}$ are
$(e^{ij})_{kl
   }= \delta_{il} \delta_{jk}$.
We also notice the identity $ R_{1,\alpha}^1(0) = \cal P $, where $\cal P $ is
the permutation operator ${\cal P} V\otimes h_\alpha = h_\alpha \otimes V $.

The $GL(N)$ mixed system is defined in terms of its
transfer matrix of alternating isomorphic~($k=1$) and
non-isomorphic vertex operators by
\EQ
T_{1,k}(\mu)= Tr_{V^{(1)}}[ \tau_{1,k}(\mu)],~~~ \tau_{1,k}(\mu)=
R_{1,L}^1(\mu)
R_{k,L-1}^1(\mu) \cdots R_{1,2}^1(\mu)
R_{k,1}^1(\mu)
\label{T1k}
\EN
where the matrices product and the trace are defined in the auxiliary space
$V^{
   (1)}$ of the elements $e^{ij}$ and
$\tau_{1,k}(\mu)$ is the so-called monodromy matrix. The associated one
dimensional
quantum Hamiltonian acting on the Hilbert
space $\prod_{\alpha=1}^{L} h_{\alpha}$ is determined as the logarithm
derivative of the transfer matrix, $H_{1,k}= J \frac{d}{d \mu}
[\ln(T_{1,k}(\mu)
   ]|_{\mu=0}$. By using this relation
and some properties of the $GL(N)$ group we find
\bear
H_{1,k}= \frac {\tilde{J}}{2} \sum_{n=odd}^{L-1} \left [ \sum_{i,j,l,m}^{N}
e^{i
   j}_{n-1} \{ E^{jm}_n,
E^{li}_n \} e^{ml}_{n+1} - (k-1) \sum_{i,j,l}^{N} ( e^{ij}_{n-1} E^{jl}_n
e^{li}
   _{n+1}  +
e^{il}_{n-1} E^{ji}_n e^{lj}_{n+1} ) \right .  \nonumber \\  \left . +
\frac{(k-1)^2}{2} \sum_{i,j}^{N}e^{ij}_{n-1} e^{ji}_{n+1} +
\sum_{i,j}^{N} ( e^{ij}_{n-1} E^{ji}_n + E^{ij}_n e^{ji}_{n+1} ) -(k-1) \right
]
    -\frac{JL}{2}(1+ \frac 2{k+1})
\label{ham}
\ear
where $\tilde{J}=4 J/(k+1)^2$
and we have conveniently added an extra constant $ -J \frac L2 (1+ \frac
2{k+1})
   $.
In this paper we are interested in the antiferromagnetic
properties of this model, and therefore we have chosen $J=1$. Setting $k=1$ in
Eq.(\ref{ham}) we reproduce the
results of refs.\cite{SU,KR} and for $N=2$ and $k=2S$ we recover previous
calculations for the
mixed Heisenberg chain \cite{VW,AM,VN}.

The diagonalization of the transfer matrix $T_{1,k}(\mu)$~(or the one
dimensional Hamiltonian $H_{1,k}$) follows
from the generalization of the quantum inverse scattering~(QIS) approach
\cite{FT}
applied to multi state vertex models \cite{KR,BVV,KR1}.
In the QIS construction the definition of the pseudo vacuum and the block form
of the
monodromy matrix play as two important ingredients of this method. In our case
the pseudo vacuum $|0 >$ is
defined by
\EQ
|0>= |0>_1^k \otimes |0>_2^1 \otimes \cdots \otimes |0>_{L-1}^k \otimes |0
>_L^1
\EN
where $|0 >_{\alpha}^k$ is
the vector of highest weight of the $GL(N)$ algebra at order $k$ of the
symmetric representation
acting on the site $\alpha$, namely $E^{ij}_{\alpha}|0>_{\alpha}^k=
k \delta_{i,1} \delta_{j,1} |0>_{\alpha}^k$. An important
property of this state is that $\tau_{1,k}(\mu) |0>$ has the following block
triangular form
\EQ
\tau_{1,k}(\mu) |0> =  \left( \begin{array}{cc}
(\mu+1)^{L/2}(\mu+(k+1)/2)^{L/2}

& B^i(\mu) \\ 0 & \delta_{i,j}(\mu)^{L/2}(\mu-(k-1)/2)^{L/2}
\end{array} \right) |0>
\label{tau0}
\EN
where $B^i(\mu) \equiv \tau_{1,k}(\mu)^{1,i+1}, i=1, \cdots, N-1$.

Following the QIS machinery \cite{KR,BVV,KR1} a certain linear combination of
the states
$ B^{j_1}(\mu_1^1) \cdots B^{j_n}(\mu_n^1) |0 >$
can be considered as a basis for the eigenstates of the transfer matrix,
provided that the parameters $\{ \mu_1^1, \cdots, \mu_n^1 \}$ satisfy
a set of non-linear equations denominated
Bethe ansatz equations.  The technical steps are fairly parallel with those of
refs. \cite{KR,BVV,KR1} and
in appendix A we have collected some of the details. Defining
the convenient shift $\mu_j^r= i\lambda_j^r -r/2$, the
Bethe ansatz equations are given by
\begin{eqnarray}
{\left ( \frac{\lambda_j^1 -i/2}{\lambda_j^1 +i/2}\right )}^{L/2}
{\left ( \frac{\lambda_j^1 -ik/2}{\lambda_j^1 +ik/2}\right )}^{L/2}
=
-\prod_{l=1}^{M^1} \frac{\lambda_j^1 -\lambda_l^1 -i}{\lambda_j^1-\lambda^1 +i}
\prod_{l=1}^{M^2} \frac{\lambda_j^1 -\lambda_l^2 +i/2}{\lambda_j^1-\lambda^2
-i/
   2} \nonumber \\
\prod_{l=1}^{M^r} \frac{\lambda_j^r -\lambda_l^r -i}{\lambda_j^r-\lambda^r +i}
   =
\prod_{l=1}^{M^{r+1}} \frac{\lambda_j^{r+1} -\lambda_l^{r+1}
-i/2}{\lambda_j^{r+
   1}-\lambda^{r+1} +i/2}
\prod_{l=1}^{M^{r-1}} \frac{\lambda_j^{r-1} -\lambda_l^{r-1}
-i/2}{\lambda_j^{r-
   1}-\lambda^{r-1} +i/2},
\label{bae}
\end{eqnarray}
where $r=2, \cdots, N-1$ and we define $M^N \equiv 0$. The eigenvalues
of the Hamiltonian $H_{1,k}$ are parametrized by
\EQ
E_{1,k}= -\sum_{j=1}^{M^1} \frac{1}{(\lambda_j^1)^2 +(1/2)^2}
\label{E1k}
\EN

In principle, the same construction discussed above can be carried out for the
`
   `dual'' one dimensional
$H_{k,1}$ spin chain. The main task is to find an isomorphic  $GL(N)$ invariant
   operator $R_{k,\alpha}^k(\mu)$
in the symmetric representation of order $k$. Using the
approach of ref. \cite{KRS}, Johannesson \cite{HO} has
explicitly exhibited such operator. Hence, analogously to Eq.(\ref{T1k}),
the associated transfer matrix $T_{k,1}(\mu)$ is
expressed by
\EQ
T_{k,1}(\mu)= Tr_{V^{(k)}}[ \tau_{k,1}(\mu)],~~~ \tau_{k,1}(\mu)=
R_{1,L}^k(\mu)
R_{k,L-1}^k(\mu) \cdots R_{1,2}^k(\mu)
R_{k,1}^k(\mu)
\label{Tk1}
\EN
 where $V^{(k)}$ is the space of the matrices $E^{ij}$ of $GL(N)$ at order $k$
of representation.

An important property of $T_{1,k}(\mu)$ and $T_{k,1}(\mu^{'})$ is their
commutativity for arbitrary values of the
parameters $\mu$ and $\mu^{'}$. This property
follows from the ``mixed'' Yang-Baxter relation satisfyed by the non-local
vertices\footnote{The vertex $R_k^{k^{'}}(\mu)$ acts in the non
local space $V^{(k^{'})} \otimes h$.}
$R_k^k(\mu)$ and $R_k^1(\mu)$. As a consequence,
the Hamiltonians $H_{1,k}$ and $H_{k,1}$
can be  simultaneously diagonalized and their eigenspectrum are
parametrized by the same Bethe equations, i.e., Eqs.(\ref{bae}).
However, the eigenenergies of $H_{k,1}$ \cite{HO} are expressed in terms of the
   variables
$\{\lambda_j^1 \}$ by a different function of that of Eq.(\ref{E1k}), namely
\EQ
E_{k,1}= - \sum_{j=1}^{M^1} \frac{k}{(\lambda_j^1)^2 +(k/2)^2}
\label{Ek1}
\EN

At this point, it is important to remark that  the transfer matrix
$T_{1,k}(\mu)
   $ and
its ``dual'' $T_{k,1}(\mu)$ are
not rotational invariant in the horizontal/vertical space of states. A
simple way of defining \cite{VW} a ``mixed''  symmetric transfer matrix
$T^{sym}
   (\mu)$,
preserving the rotational invariance,
 is by formally multiplying
these two transfer matrices
\EQ
T^{sym}(\mu)= T_{1,k}(\mu) T_{k,1}(\mu)
\label{Tsym}
\EN

Due to the commutativity between $T_{1,k}(\mu)$ and $T_{k,1}(\mu)$ the spectrum
   of $T^{sym}(\mu)$ is clearly
parametrized by the same Bethe ansatz equation and the eigenvalues of the
corresponding
one dimensional Hamiltonian $H^{sym}$ are  added,
\EQ
E^{sym}=
 - \sum_{j=1}^{M^1} \frac{1}{(\lambda_j^1)^2 +(1/2)^2}
 - \sum_{j=1}^{M^1} \frac{k}{(\lambda_j^1)^2 +(k/2)^2}
\label{Esym}
\EN

 In this sense Eqs.(\ref{bae}, \ref{E1k}, \ref{Ek1}, \ref{Esym} )
define three families of possible
spectrums parametrized by a single Bethe ansatz
equation. For instance, in the case of small size $L$, one only needs
to solve Eq.(\ref{bae}) and compare
it with the exact solution of the spectrum of $H_{1,k}$ in order
to investigate the structure of the
variables $\{\lambda_j^r \}$ which are going to parametrize the
whole spectrum of all
these three families of models. In general, we
remark that a certain solution $\{\lambda_j^r\}$ of Eq.(\ref{bae}) will not
necessarily produce the same $i^{th}$ state
ordered in energy for all these models. The ground state, however,
is characterized by the same structure of $\{\lambda_j^r \}$ for all
these three families. In the
thermodynamic limit, $L \rightarrow \infty$, the ground state
solution $\{ \lambda_j^r \}$ is composed
by a mixture of real numbers $\eta_j^r $ and a set of complex
roots $\lambda_{j,\alpha}^{r}$. The complex structure
$\lambda_{j,\alpha}^{r}$ cluster in the so-called k-string form
\EQ
\lambda_{j,\alpha}^{r}= \xi_j^r +\frac{i}{2}(k+1-2\alpha),~~~
\alpha=1,2,\cdots,
   k
\label{kstring}
\EN
where $\xi_j^r$ is a real parameter denominated center of the k-string.

 In order to calculate the ground state energy in the $L \rightarrow \infty$
limit, one has to
solve the Bethe ansatz equations (\ref{bae}) for the variables $\eta_j^r$ and
$\
   xi_j^r$. Taking into
account Eq.(\ref{kstring}), and after some standard manipulations we are able
to
    compute the following
values for the ground state energy per particle
\begin{eqnarray}
e_{\infty}^{1,k} &=& -\frac{1}{N} \left [ \psi(1) -\psi(1/N)
+\psi(\frac{k-1+2N}
   {2N})
-\psi(\frac{k+1}{2N}) \right ] \\
e_{\infty}^{k,1} &=& -\frac{1}{N} \left [ \psi(1) -\psi(k/N)
+\psi(\frac{k-1+2N}
   {2N})
-\psi(\frac{k+1}{2N}) +\sum_{j=1}^{k-1}\frac{N}{j} \right ]
\end{eqnarray}
where $\psi(x)$ is the Euler psi function. From Eq.(12)
it is clear that $e_{\infty}^{sym}=e_{\infty}^{1,k} +e_{\infty}^{k,1}$.

Let us now concentrate our attention on the low-lying excitations of the
symmetric mixed model.
Since the transfer-matrix $T^{sym}(\mu)$ is rotational invariant, its
associated
    quantum spin Hamiltonian
is a strong candidate to be conformally invariant. We recall that the analysis
of the critical properties
in spin chains depends on the behaviour of its dispersion relation for low
momenta $p$. The computation of the dispersion relation follows the standard
formalism of perturbing the ground
state structure by holes
and string of arbitrary length~(see e.g. \cite{SU}). The only
subtle fact is that this alternating mixed system the total momenta is half of
those
considered in homogeneous models~($k=1$) \cite{AM}. Following refs.
\cite{SU,HO,
   VW} we find that the
dispersion relation for all branches of excitations \cite{SU,HO} possess the
same linear behaviour
for the total low momenta $p$
\EQ
\varepsilon(p) = \frac{4 \pi}{N} p
\label{epsp}
\EN

{}From Eq.(\ref{epsp}) the sound velocity is $v_s= 4 \pi/N$, independent of the
order $k$ of representation.
We observe that $v_s$ is double of that appearing in the homogeneous
model~($k=1
   $) \cite{SU,HO}. This
fact can be easily interpreted, by noticing that the elementary translation
``cell'' of the alternating
mixed models is double of that  of homogeneous system. Indeed, considering
this discussion and the previous results
of ref. \cite{SU,HO}~(assuming independence of $k$) will
lead us to guess that $v_s= 4 \pi/N$, with no need of an explicit computation
\footnote{
In fact, for an alternating  model with periodicity $l$~( made by a
collection of ``spins'' operators at different order of representation), we
should have $v_s=2l \pi/N$.}.

In the next section we are going to compute the conformal anomaly which defines
   the class of
universality of these conformal mixed $GL(N)$ models.

\section{The thermodynamics of the mixed GL(N) model}

In order to discuss the thermodynamic properties we adopt the thermodynamic
Bethe ansatz approach
originally proposed by Yang and Yang \cite{YY}. This method is based on the
minimization of the free energy
and takes advantage of the integrability through the Bethe ansatz equations.
The
    first
step is to notice that the Bethe ansatz equations (\ref{bae}) admit the same
string hypothesis used
previously by Takahashi \cite{TA} in the isotropic Heisenberg model. This
observation allows us to conclude
that , in the $L \rightarrow \infty$ limit, the parameters $\lambda_j^r$
organized in strings of the type
described in
Eq(\ref{epsp}). Substituting Eq.(\ref{epsp}) in Eq.(\ref{bae}) and
taking the thermodynamic limit, we are able to obtain the
following infinity set
of  coupled integral equations for the densities
$\sigma_n^r(\lambda)$~($\tilde{\sigma}_n^r(\lambda))$of particles~(holes)
\EQ
\tilde{\sigma}_n^r(\lambda)= \frac{\delta_{r,1}}{4 \pi} \left [
\phi_{n,1/2}(\lambda) +\phi_{n,k/2}(\lambda)
\right ] -\sum_{r^{'}=1}^{N-1} \sum_{j=1}^{\infty}
(A_{n,j}*B_{r,r^{'}}*\sigma_j
   ^{r^{'}})(\lambda)
\EN
where $n$ indicates the length of the n-string, and $(f*g)(x)$ denotes the
convolution $\frac{1}{2 \pi}
\int_{-\infty}^{\infty} f(x-y)g(y) dy$. The functions $A_{n,j}(\lambda)$,
$B_{r,
   r^{'}}(\lambda)$ and
$\phi_{n,j}(\lambda)$ are easily represented in terms of their Fourier
transforms. Defining the Fourier
component of a given function $f(x)$ by $f(\omega)= \frac{1}{2 \pi}
\int_{\infty
   }^{\infty} dx e^{-i \omega x}f(x)$,
we have the following expressions for $A_{n,j}(\omega)$, $B_{r,r^{'}}(\omega)$
and $\phi_{n,j}(\omega)$
\begin{eqnarray}
A_{n,j}(\omega)&=& \coth(|\omega|/2) \left [ e^{-|n-j||\omega|/2}
-e^{-(n+j)|\omega|/2} \right ] \\
B_{r,r^{'}}(\omega) &=& \delta_{r,r^{'}} -p(\omega) l_{r,r^{'}} \\
\phi_{n,j/2}(\omega) &=& A_{n,j}(\omega) p(\omega)
\end{eqnarray}
where  $p(\omega)= \frac{1}{2 \cosh(\omega/2)}$ and $l_{r,r^{'}}$ is the
incident matrix of the $A_{N-1}$
Lie algebra.

The second step is to encode the temperature $T$ via minimization of the free
energy $F^{sym}=E^{sym}-TS$.
By using a standard procedure \cite{YY,TA,BA,TW} the energy $E^{sym}$ and the
entropy $S$ can be
written in terms of the densities of particles~($\sigma_n^r(\lambda)$)
and holes~($\tilde{\sigma}_n^r(\lambda)$).
After the minimization, $ \delta F^{sym}=0$, we get the following thermodynamic
   Bethe ansatz~(TBA) equations
\begin{eqnarray}
\epsilon_n^r(\lambda)&=& K_{r}(\lambda)( \delta_{n,1} +\delta_{n,k}) +
T \sum_{r^{'}}^{N-1} \varphi_{r,r^{'}}* \left [ \ln(1+
e^{\epsilon_n^{r^{'}}/T})
\right ](\lambda) \label{TBA}\nonumber \\
                     & & T \sum_{r^{'}}^{N-1} \tilde{\varphi}_{r,r^{'}}* \left
[
    \ln(1+ e^{\epsilon_{n+1}^{r^{'}}/T}) +
 \ln(1+ e^{\epsilon_{n-1}^{r^{'}}/T}) \right ](\lambda)
\end{eqnarray}
where
$\tilde{\sigma}_n^r(\lambda)/\sigma_n^r(\lambda)=e^{\epsilon_n^r(\lambda)
   /T}$, $\epsilon_0^r(\lambda)
\equiv 0 $, and the Fourier component of the functions $K_r(\omega)$,
$\varphi_{
   r,r^{'}}(\omega)$ and
$\tilde{\varphi}_{r,r^{'}}(\lambda)$ are given by
\begin{eqnarray}
K_{r}(\omega) &=& p(\omega) B_{r,1}^{-1}(\omega) \\
\varphi_{r,r^{'}}(\omega) &=& \delta_{r,r^{'}} -B_{r,r^{'}}^{-1},~~~
\tilde{\varphi}_{r,r^{'}}(\omega)=p(\omega)
B_{r,r^{'}}^{-1}
\end{eqnarray}

Finally, the equilibruim free energy $F^{sym}$ is given by
\EQ
F^{sym}/L = e_{\infty}^{sym} -\frac{T}{4 \pi} \int_{-\infty}^{\infty} d \lambda
   \sum_{r=1}^{N-1} K_{r}(\lambda)
\left [ \ln( 1+e^{\epsilon_1^r(\lambda)/T}) + \ln(1+
e^{\epsilon_k^r(\lambda)/T}
   ) \right ]
\label{FreeE}
\EN

One possible way to calculate the central charge of a
conformally invariant system is by analysing the low temperature
behaviour of the respective free energy. The universal behaviour of the free
ene
   rgy is given by \cite{CA1,AF}
\EQ
F/L= e_{\infty} -\frac{\pi c T^2}{ 6 v_s }
\label{FoverL}
\EN

It turns out that Eqs(\ref{TBA}, \ref{FreeE}) allow us to make an exact
calculation of
such low temperature behaviour. We first define the
shift $\lambda \rightarrow \lambda -\frac{N}{2 \pi} \ln(\frac{N T}{2 \pi})$,
taking the derivative in $\lambda$ of
Eq.(\ref{TBA}) and after some few standard manipulations \cite{BA} the $T
\rightarrow 0$ limit can be expressed
in terms of the Dilogarithm functions $L(x)$ by
\EQ
F^{sym}/L = e_{\infty}^{sym} -\frac{N T^2}{24} \left [ (N-1)k -\sum_{r=1}^{N-1}
   \sum_{m=1}^{k-2}
L(\frac{\sin[(k-m-1)\theta] \sin[m \theta]}{\sin[(m+r)\theta]
\sin[(r+k-m-1)\theta]}) \right ]
\EN
where $\theta= \frac{\pi}{N+k-1}$ and $L(x)= -\frac{3}{\pi^2} \int_{0}^{x} dt [
   \frac{\ln(t)}{1-t} +
\frac{\ln(1-t)}{t} ]$.

Using some identities for the sum of the Dilogarithm function proved in ref.
\cite{KI}, we finally have
\EQ
F^{sym}/L = e_{\infty}^{sym} -\frac{T^2}{24} \frac{(N-1)((N+2)k-2)}{N+k-1}
\label{FoverL2}
\EN

 Comparing Eqs.(\ref{FoverL}, \ref{FoverL2}), we find
that the central charge is $c=(N-1)((N+2)k-2)/(N+k-1)$.
Remarkably enough, this conformal anomaly can be
decomposed in terms of the central charges of two $SU(N)$
WZWN models with topological charge $\tilde{k}=1$~($c=N-1$) and
$\tilde{k}=k-1$~($c=(N^2-1)(k-1)/(N+k-1)$). This result generalizes
similar decomposition mentioned by the authors \cite{AM} for
the $SU(2)$  mixed Heisenberg model. In order
to give an extra support for this value of the central charge,
in appendix B we present some numerical
results for the finite size effects of the ground state energy.

\section{ Discussions on possible generalizations}

It is almost evident that all of our discussion of section 2 can be generalized
   to an arbitrary
representation of order $k$ in the auxiliary space of states. The main
technical
    difficulty is
the explicit construction of the non-isomorphic vertex operator
$R_{k,j}^{k^{'}}
   (\mu)$. The solution
of this problem has already been considered in ref. \cite{KRS} for arbitrary
finite representations
of $GL(2)$ Lie algebra. The vertex $R_{k,j}^{k^{'}}(\mu)$ is expressed as a
linear combination of certain
projectors defined on the subspaces of the Klebsch-Gordon decomposition of
$GL(2
   )_k \otimes GL(2)_{k^{'}}$.
The transfer matrix $T_{k,k^{'}}(\mu)$ is then defined by
\EQ
T_{k,k^{'}}(\mu)= Tr_{V^{(k)}}[ R_{k,L}^k(\mu)
R_{k^{'},L-1}^{k}(\mu) \cdots R_{k,2}^k(\mu)
R_{k^{'},1}^{k}(\mu) ]
\EN

 The eigenvectors and the eigenvalues of the associated one dimensional
Hamiltonian $H_{k,k^{'}}$ can be
determined by using the following strategy. We first define
an auxiliary transfer matrix $T_{k,k^{'}}^{aux}(\mu)$
which commutes with $T_{k,k^{'}}(\mu)$ as
\EQ
T_{k,k^{'}}^{aux}(\mu)= Tr_{V^{(1)}}[ R_{k,L}^1(\mu)
R_{k^{'},L-1}^1(\mu) \cdots R_{k,2}^1(\mu)
R_{k^{'},1}^1(\mu) ]
\label{Tkkl}
\EN

 This definition has the advantage of reducing the auxiliary space to its
fundamental representation, hence
the standard QIS formalism can be applied. On the other hand, the eigenvalues
of
    $T_{k,k^{'}}(\mu)$
can be related to those of $T_{k,k^{'}}^{aux}(\mu)$ by  certain commutators of
$
   T_{k,k^{'}}(\mu)$
and the usual quantum
inverse scattering $B(\mu)$ operator \cite{BA}. The eigenvalues of the
$H_{k,k^{
   '}}$ are parametrized
by the Bethe ansatz equation
\EQ
{\left ( \frac{\lambda_j -ik/2}{\lambda_j +ik/2}\right )}^{L/2}
{\left ( \frac{\lambda_j -ik^{'}/2}{\lambda_j +ik^{'}/2}\right )}^{L/2}
=
-\prod_{l=1}^{M} \frac{\lambda_j -\lambda_l -i}
{\lambda_j -\lambda_l +i}
\label{bae2}
\EN
where $\mu_j=i\lambda_j-1/2$ and the eigenvalues $E_{k,k^{'}}$ are determined
by
\EQ
E_{k,k^{'}}= - \sum_{j=1}^{M} \frac{k}{(\lambda_j)^2 +(k/2)^2}
\label{Ekkl}
\EN

Analogously to what we have discussed in section 2,  similar approach works for
   $T_{k^{'},k}(\mu)$ and the
conformally invariant quantum Hamiltonian can be defined through
the product $T_{k,k^{'}}(\mu) T_{k^{'},k}(\mu)$.
Considering the results of section 2 and those of ref. \cite{KRS} on the
Yang-Baxter solutions
for the $GL(N)$ group, it seems plausible that similar conclusions reached for
$
   N=2$ can be extended for
an arbitrary value of $N$. Comparing  the left hand
side of Eqs(\ref{bae}, \ref{bae2}) in the case of $N=2$, we observe that a
certain factor
$1/2$ has been replaced by $k/2$. This leads us to
conjecture that similar mechanism should work for a general mixed $GL(N)$
system. Taking into account this fact and writing Eq.(\ref{bae}) in a more
convenient way, we conjecture that the
 form of a $GL(N)_k \otimes GL(N)_{k^{'}}$  Bethe ansatz equation is
\EQ
{\left ( \frac{\lambda_j^r -i\delta_{r,1}k/2}{\lambda_j^r
+i\delta_{r,1}k/2}\right )}^{L/2}
{\left ( \frac{\lambda_j^r -i\delta_{r,1} k^{'}/2}{\lambda_j^r +i\delta_{r,1}
k^
   {'}/2}\right )}^{L/2}
=
-\prod_{r^{'}=1}^{N-1}\prod_{l=1}^{M^r} \frac{\lambda_j^{r^{'}}
-\lambda_l^{r^{'
   }} -iC_{r,r^{'}}/2}
{\lambda_j^{r^{'}} -\lambda_l^{r^{'}} +iC_{r,r^{'}}/2}
\label{baeCrr}
\EN
where $C_{r,r^{'}}$ is the $A_{N-1}$ Cartan matrix
and the eigenvalues $E^{sym}$ of the conformally invariant Hamiltonian are
given
    by
\EQ
E^{sym}= - \sum_{j=1}^{M^1} \frac{k}{(\lambda_j^1)^2 +(k/2)^2}
 - \sum_{j=1}^{M^1} \frac{k^{'}}{(\lambda_j^1)^2 +(k^{'}/2)^2}
\label{Ekksym}
\EN

It is not a surprise that the structure of the Bethe ansatz equation is closely
   related to the
$A_{N-1}$ Lie algebra. In the case of homogeneous vertex models~($k=k^{'}$),
the

authors of ref.\cite{CON}
have conjectured that the same structure will remain for all semi-simple A,D,E
Lie algebras\footnote{
This fact is related to the idea that the classification of the solutions of
the Yang-Baxter equations is somewhat connected to the classification of the
Lie
    algebras and their
automorphisms.}. In fact in ref. \cite{ON} this conjecture has been  verified
by
    an explicit computation
in the case of $D_n$ Lie algebra. Based on these observations, let us assume
that the same
conjecture can be extented to the case of non-homogeneous~($k \otimes k^{'}$)
models. It is not difficult
to verify that the associated TBA equations are similar to the
system of Eqs.(\ref{TBA}). We just have to
replace $l_{r,r^{'}}$ and $N$ in Eqs.(\ref{TBA}) by the incident matrix and the
   rank of the
corresponding A,D,E Lie algebra. Interesting enough, these equations can be
cast
    in a rather useful form
which will be helpful in the analysis of the low temperature . Defining the
function $Y_n^r(\lambda)=
e^{-\epsilon_n^r(\lambda)/T}$, making few manipulations in the Fourier
transform
    of Eq.(\ref{TBA}) and Fourier
transforming back we find the following expression
\EQ
Y_n^r(\lambda+i/2)Y_n^r(\lambda-i/2)
\prod_{r^{'} \in G}{\left [ 1+Y_n^{r^{'}}(\lambda) \right]}^{-l_{r,r^{'}}}
\prod_{j \in A_{\infty}}{\left [ 1+1/Y_j^r(\lambda) \right]}^{l_{n,m}}
=e^{2 \pi \delta(\lambda) \delta_{r,1}(\delta_{n,k} +\delta_{n,k^{'}})/T}
\label{YY}
\EN
where $r^{'}$ is an index characterizing the nodes of the Dynkin diagram of the
   $G\equiv A,D,E$ Lie algebra
and $j$ is a similar~(unrestricted) index for the $A_{\infty}$ Lie algebra.

Eq. (\ref{YY}) defines a set of functional hierarchy relations for the
functions
    $Y_n^r(\lambda)$. The
possibility of constructing such functional relations from the TBA equations
has
    been first noted
by Al. Zamolodchikov \cite{ZA} in the case of the diagonal system of scattering
   $S$-matrices. It also
appears that certain functional hierarchies play the keystone in the
computation
    of critical
dimensions in the integrable lattice models \cite{PJ}. In our case they encode
all
necessary information in order to obtain the low temperature behaviour of the
free energy. Procceding as in
sect.3 , we can show that the
$T \rightarrow 0$ limit of the free energy assumes the following form
\EQ
F^{sym}/L =e_{\infty}^{sym} -\frac{T^2 h_G}{24 \pi} \left [ k^{'} r_{G} -
\sum_{r \in G} \sum_{j \in A_k} L(1/1+y_j^r(k))
-\sum_{r \in G} \sum_{j \in A_{k^{'}-k}} L(1/1+y_j^r(k^{'}-k)) \right ]
\label{FoverL3}
\EN
where $r_{G}$ and $h_G$ are the rank and the dual Coxeter number of the Lie
algebra $G$, and
$e_{\infty}^{sym}= -\int_{-\infty}^{\infty} d \omega \frac{B_{1,1}^{-1}}{4
\cosh
   (\omega/2)}[
\phi_{k,k/2}(\omega) +\phi_{k^{'},k^{'}/2}(\omega) +2 \phi_{k^{'},k/2}(\omega)]
    $. The
constants $y_j^r(m)$ satistfy
the equation
\EQ
y_j^r(m)^2=
\prod_{r^{'} \in G} {\left [ 1+y_j^{r^{'}}(m) \right ]}^{l_{r,r^{'}}}
\prod_{i \in A_m } {\left [ 1+1/y_i^r(m) \right]}^{-l_{i,j}}
\label{yrj}
\EN
where $l_{r,r^{'}}$~($l_{i,j}$) is the incident matrix of the Lie algebra
$G$~($
   A_m$).

Remarkably enough, the sum of Dilogarithm function \cite{KI,KU}
appearing in Eq.(\ref{YY}) has been
conjectured to have the expression
\EQ
\sum_{r \in G} \sum_{j \in A_m} L(1/1+y_j^r(m))= \frac{r_G m(m-1)}{h_G +m}
\EN

Although the proof of last identity was essentially given for the $A$ and $D$
Lie algebra, it is
quite direct to verify it by numerically
solving Eq.(\ref{yrj}) for several small values of $m$ and $h_G$ \cite{KU,MA}.
Using this Dilogarithm sum and taking into account that now $v_s= 4 \pi/h_G$,
we
    finally
obtain the following central charge
\EQ
c= \frac{r_G k (h_G+1)}{h_G+k} + \frac{ r_G (k^{'}-k)(h_G+1)}{h_G +k^{'}-k}
\EN
where we have already decomposed the result in terms of the central charge of
two $G$ invariant
WZWN models with topological charges $\tilde{k}=k$ and $\tilde{k}=k^{'}-k$,
respectively. In particular for $k^{'}=k$ we recover the known conjecture that
rational isomorphic vertex models are
in the class of universality of WZWN field theories \cite{CV,AF,MA}.

\section{Conclusion}
In this paper we have discussed the critical behaviour of conformally invariant
   mixed $GL(N)$
spin chains. The central charge of an alternating model with ``spins''
operators at order $k$~($k^{'}$) of representation acting on even~(odd) sites
can be decomposed in terms of WZWN field theories
with topological charge $k$ and $k^{'}-k$~($k^{'}>k)$. We have also considered
the generalization
of this result to other symmetric
representation of the $A,D,E$ Lie algebras.

Another possible extension of the results of this paper is to consider a
collection of vertex operators
at different representation distributed on a line  of size $L$ and
with periodicity $l$. The associated rotational
symmetric transfer matrix can be defined as
\EQ
T_{k_1,\cdots,k_l}^{sym}(\mu)= \prod_{i=1}^l Tr_V^{(k_i)} [
R_{k_1,L}^{k_i}(\mu)
    \cdots
R_{k_l,L-l-1}^{k_i}(\mu) \cdots
R_{k_1,l}^{k_i}(\mu) \cdots R_{k_l,1}^{k_i}(\mu) ]
\label{TsymGenr}
\EN

Following our considerations of sections 3 and 4, the basic change in the TBA
equations (34) is that the
right hand side of Eqs.(\ref{YY}) is replaced by
$2 \pi \delta(\lambda)\delta_{r,1} \sum_{i=1}^{l} \delta_{n,k_i}/T$. For
instance,
taking the following ordering $k_1< k_2 < \cdots < k_l$, the central charge of
the  one
dimensional Hamiltonian associated to the system (\ref{TsymGenr}) will be
\EQ
c= \sum_{i=0}^{l-1} c(k_{i+1}-k_i);~~~ k_0 \equiv 0
\label{central}
\EN
where $c(k)$ is the central charge of the $A,D,E$ WZWN model with topological
charge $k$.

Finally, it should be interesting to study the full operator content of these
models and in particular
to understand the decomposition of Eq.(\ref{central}) in terms of the bosonic
and parafermionic fields of the WZWN theories.
We notice that for the sequence $k_{i+1}-k_i=1$ and $k_1=1$, the central charge
   is $l$ time the rank of $G$ and hopefully
the operator content will be determined by $lr_G$ coupled bosonic fields.

\section*{Acknowledgements}
The work of M.J. Martins was partially supported
by CNPq~(Brazilian agency).\\
\vspace{1.0cm}\\
\centerline{\bf Appendix A}
\setcounter{equation}{0}
\renewcommand{\theequation}{A.\arabic{equation}}

Following the basic steps of the quantum inverse scattering method we propose
a set of eigenstates $|\psi>$ defined by \cite{YB1,FT,BVV}
\EQ
|\psi> \equiv \psi(\mu_1^1,\cdots,\mu_n^1)=
F_{j_1 \cdots j_{M^1}} B^{j_1}(\mu_1^1) \cdots B^{j_{M^1}}(\mu_{M^1}^1) |0 >
\label{psi}
\EN
where $|0>$ is the pseudo vacuum~(see Eq.(5)). The next step,
motived by the properties of $\tau_{1,k}(\mu)|0>$,
is to decompose the monodromy matrix as
\EQ
\tau_{1,k}(\mu) =  \left( \begin{array}{cc} A(\mu) & B^i(\mu) \\ C^j(\mu) &
D^{i
   j}(\mu)
\end{array} \right)
\label{tau}
\EN
where $i(j)$ is a row(column) index, $i,j=1, \cdots, N-1$.

It follows from the identity $T_{1,k}(\mu)=Tr_{V^{(1)}} [\tau_{1,k}(\mu)]$ that
\EQ
T_{1,k}(\mu) |\psi> = [A(\mu) + \sum_i D^{ii}(\mu)] |\psi>
= \Lambda(\mu,\{\mu_i^1\}) |\psi>
\label{Tpsi}
\EN

In order to find the eigenstates $|\psi>$ and the eigenvalues $
\Lambda(\mu,\{\mu_i^1\})$ we also need
the commutation relation between the $A(\mu)$, $B^i(\mu)$ and $D^{ii}(\mu)$
operators. These relations
follows from the relation
\EQ
\tilde{R}_1^1(\mu^{'}-\mu) \tau_{1,k}(\mu^{'}) \otimes \tau_{1,k}(\mu ) =
\tau_{1,k}(\mu) \otimes \tau_{1,k}(\mu^{'} ) \tilde{R}_1^1(\mu^{'}-\mu)
\label{YB}
\EN
where $\tilde{R}_1^1(\mu)={\cal P} R_1^1(\mu)$.  Using
Eqs.(\ref{R1k},\ref{tau},
   \ref{YB})
we have
\bear
A(\mu^{'}) B^i(\mu) = \frac{\mu-\mu^{'}+1}{\mu-\mu^{'}} B^i(\mu) A(\mu^{'}) +
\frac 1{\mu^{'}-\mu} B^i(\mu^{'}) A(\mu) \nonumber \\
D^{ii}(\mu^{'}) B^k(\mu) = \frac 1{\mu-\mu^{'}} B^k(\mu^{'}) D^{ii}(\mu) +
\frac 1{\mu^{'}-\mu} \sum B^k(\mu) D^{ii}(\mu^{'})
[R_1^1(\mu^{'}-\mu)]^{lk,ni}_
   {[N-1]}
\label{comut}
\ear
where $[R_1^1(\mu^{'}-\mu)]^{lk,ni}_{[N-1]}$  are the $GL(N-1)$ matrix elements
of the matrix defined in Eq.(\ref{R1k}).

{}From Eqs.(\ref{psi},\ref{YB},\ref{comut}) follows that
\bear
A(\mu)|\psi> &=& \prod_{i=1}^{M^1} \frac{\mu_i^1-\mu+1}{\mu_i^1-\mu}
a(\mu) |\psi> + \mbox{U.T.} \nonumber \\
D^{ii}(\mu)|\psi> &=&(\prod_{i=1}^{M^1} \frac 1{\mu-\mu_i^1})
{t^{(2)}}^{i_1 \cdots i_{M^1}}_{l_1 \cdots l_{M^1}}
F^{i_1 \cdots i_{M^1}} B^{l_1}(\mu_1^1) \cdots B^{l_{M^1}}(\mu_{M^1}^1)
d(\mu) |\psi> + \mbox{U.T.} \nonumber \\
\label{ADs}
\ear
where $a(\mu) = (\mu+1)^{L/2}(\mu+\frac 12 (k+1)^{L/2}$,
$d(\mu) = \mu^{L/2}(\mu-\frac 12 (k-1))^{L/2}$ and
${t^{(2)}}^{i_1 \cdots i_{M^1}}_{l_1 \cdots l_{M^1}}$
are matrix elements of the following operator
\EQ
t^{(2)}(\mu;\{\mu_j^1\}) = tr [R_{1,M^{1}}^1(\mu-\mu_{M^1}^1)_{[N-1]} \cdots
R_
   {1,1}^1(\mu-\mu_1^1)_{[N-1}]]
\label{t11}
\EN
and $ F^{i_1 \cdots i_{M^1}}$ are eigenvector's component of
$t^{(2)}(\mu)$ with eigenvalues $\Lambda^{(2)}(\mu,\{ \mu_i^1 \})$.

The U.T. symbol stands for ``unwanted terms'' which appears due
to the interchange of the arguments $\mu$ and
$\mu^{'}$ in the relation (\ref{comut}). When these terms are null,
$|\psi>$ becomes
an eigenstate of $T_{1,k}(\mu)$ and as a consequence we obtain
a restriction to the
rapidities $\mu$~(the Bethe ansatz equation). Finally,
Eq.(\ref{ADs}) is solved by introducing in each step
$i=2, \cdots, N-1$ a new matrix $t^{(i)}(\mu)$ acting on $M^{(i)}$
sites, analogously to that of Eq.(\ref{t11}).
The final result for the eigenvalues $\Lambda(\mu)$ of $T_{1,k}(\mu)$
and $\Lambda^{(r)}(\mu,\{ \mu_i^{r-1} \})$ of
$t^{(r)}(\mu)$ are
\bear
\Lambda(\mu) & = & a(\mu) \prod_{i=1}^{M^1} \frac{\mu_i^1-\mu+1}{\mu_i^1-\mu} +
d(\mu) \prod_{i=1}^{M^1} \frac 1{\mu-\mu_i^1}\Lambda^{(2)}(\mu;\{\mu^{(1)}_j\})
 \\
\Lambda^{(r)}(\mu;\{\mu^{r-1}_j\}) & = &
\prod_{i=1}^{M^{r-1}} (\mu-\mu^{r-1}_i+1)
\prod_{i=1}^{M^r} \frac{\mu^r_i-\mu+1}{\mu^r_i-\mu} +
\prod_{i=1}^{M^{r-1}} (\mu-\mu^{r-1}_i)
\prod_{i=1}^{M^r} \frac 1{\mu-\mu^r_i} \Lambda^{(r+1)}(\mu;\{\mu^r_j\})
\nonumber
\label{eigens}
\ear

Eq.(7) is then obtained by imposing the zero residue condition in Eq.(A.8).

\vspace{1.0cm}

\centerline{\bf Appendix B}

\vspace{0.5cm}

\setcounter{equation}{0}
\renewcommand{\theequation}{B.\arabic{equation}}

The critical behaviour of a conformally invariant theory can be determined by
studying
the consequences of the finite size $L$ effects for the eigenspectrum
\cite{CA}.
    For example, the central charge is related
to the ground state energy $E^{sym}(L)$ by \cite{CA1,AF}
\EQ
E^{sym}(L)/L = e_{\infty}^{sym} -\frac{ \pi v_s c}{ L^2}
\label{eq1}
\EN

The central charge $c$ can be numerically calculated by extrapolating the
sequence
\EQ
c(L) = - [E^{sym}(L)/L -e_{\infty}^{sym}] \frac{L^2}{ \pi v_s}
\label{eq2}
\EN

In table 1(a,b) we present our estimatives for the sequence (B.2)
in the case of $N=2,3$ ( The $N=2$ data has been already
presented by us in ref.\cite{AM}.) and $k=3$.
We notice that these numerical results
are in concordance with the TBA analysis of section 3. (Eq.(27)). In our
numerical analysis
we have also observed that the case $k=2$ is rather special. The string
hypothes
   is~($\lambda_j^r=
\xi_j^r \pm i/2$) is almost exact for large enough $L$, presenting a very
unusual small
correction \cite{VW,AM}. In this case we can use the analytical method
of ref. \cite{VW1} and conclude that the central charge is $c=2(N-1)$~(in
agreement with
Eq.(27)).

\vspace{2.cm}

{\bf Table 1(a,b)} The estimatives of the central charge of Eq.(B.2) for $k=3$,
   (a) $N=2$ and (b) $N=3$.
\vspace{1.5cm}\\
{\bf (a)}\\
\vspace{0.15cm}
\begin{tabular}{|l|l|} \hline
L & $ N=2, k=3$ \\ \hline \hline
8 & 2.839 364 \\ \hline
16 &  2.602 543 \\ \hline
24 &  2.556 301 \\ \hline
32 &  2.538 530 \\ \hline
40 &  2.529 503 \\ \hline
48 &  2.524 142 \\ \hline
Extrapolated &  2.500(6) \\ \hline
\end{tabular}\\

\vspace{-7.4cm}
\hspace{6.0cm}
{\bf (b)}\\
\vspace{-6.0cm}
\hspace{6.0cm}
\begin{tabular}{|l|l|} \hline
L & $ N=3, k=3$ \\ \hline \hline
12  & 5.614 431 \\ \hline
24&  5.336 131 \\ \hline
36 &  5.277 662 \\ \hline
48 & 5.254 475 \\ \hline
60 &  5.242 437 \\ \hline
72 &  5.235 167 \\ \hline
Extrapolated &  5.206(1) \\ \hline
\end{tabular}

\end{document}